\documentclass[10pt,superscriptaddress,preprintnumbers,amsmath,amssymb,nofootinbib]{revtex4}
\usepackage{epsfig}  
\usepackage{graphicx}
\usepackage{hyperref}
\usepackage{color}
\usepackage{float}
\usepackage{amsfonts}
\usepackage{amsmath}
\usepackage{slashed}
\usepackage{soul}

\large

\begin{document} 
\title{Resolution of $R_D$/$R_{D^*}$ puzzle}
%Identifying Lorentz structure of new physics in $b\rightarrow c\tau\bar{\nu}$ transition}

\author{Ashutosh Kumar Alok}
\email{akalok@iitj.ac.in}
\affiliation{Indian Institute of Technology Jodhpur, Jodhpur 342011, India}

\author{Dinesh Kumar}
\email{dinesh@uniraj.ac.in}
\affiliation{Department of Physics, University of Rajasthan, Jaipur 302004, India}

\author{Suman Kumbhakar}
\email{suman@phy.iitb.ac.in}
\affiliation{Indian Institute of Technology Bombay, Mumbai 400076, India}

\author{S Uma Sankar}
\email{uma@phy.iitb.ac.in}
\affiliation{Indian Institute of Technology Bombay, Mumbai 400076, India}

\date{\today} 

\preprint{}

\begin{abstract}
One of the exciting results in flavor physics in recent times is the $R_D$/$R_{D^*}$ puzzle. The measurements of these flavor ratios performed by the B-factory experiments, BaBar and Belle, and the LHCb experiment are about $4\sigma$ away from the Standard Model expectation. These measurements indicate that the mechanism of $b\rightarrow c\tau\bar{\nu}$ decay is not identical to  that of $b\rightarrow c(\mu/e)\bar{\nu}$. This charge lepton universality violation is particularly intriguing because these decays occur at tree level in the Standard Model. In particular, we expect a moderately large new physics contribution to $b\rightarrow c\tau\bar{\nu}$. The different types of new physics amplitudes, which can explain the $R_D$/$R_{D^*}$ puzzle, have been identified previously. In this letter, we show that the polarization fractions of $\tau$ and $D^*$ and the angular asymmetries $A_{FB}$ and $A_{LT}$ in $B\rightarrow D^*\tau\bar{\nu}$ decay have the capability to uniquely identify the Lorentz structure of the new physics. A measurement of these four observables will lead to such an identification.
\end{abstract}
 
\maketitle
%%%%%%%%%%%%%%%%%%%%%%%%%%
\section{Introduction}
%%%%%%%%%%%%%%%%%%%%%%%%%%%%
Precision measurements in the flavor sector, both leptonic as well as hadronic, have lead to a number of discoveries in particle physics. Some of the examples are the following: 
\begin{itemize}
\item Construction of $(V-A)$ theory due to the smallness of $\pi^-\rightarrow e^-\bar{\nu}$.
\item Prediction of second neutrino due to non-occurrence of $\mu\rightarrow e\gamma$ and similarly for the prediction of the third neutrino.
\item Prediction of charm quark due to the tiny value of $K_L-K_S$ mass difference.
\item Prediction of third generation due to the discovery of CP violation.
\end{itemize}
Therefore, precision studies of flavor decays play an important role in particle physics. These studies mainly concentrated on flavor changing neutral interaction (FCNI) in meson decays. In Standard Model (SM), FCNI occur only at loop level and hence are predicted to be small. It was expected that a precise measurement of FCNI would reveal possible deviations from the SM. The theoretical predictions for these decays tend to have large uncertainties because of hadronic form factors. In recent times, a number of observables were defined for which the form factor dependence is quite weak. One such observable, $P'_5$ in the decay $B\rightarrow K^*\mu\mu$, is measured by the LHCb experiment~\cite{Aaij:2013qta,Aaij:2015oid} and is found to differ from its SM prediction by $\sim 4\sigma$ \cite{Descotes-Genon:2013vna}. Very recently, LHCb experiment also observed charged lepton universality violation in $B\rightarrow (K,\,K^*)\,l^+\,l^-$ ($l=\mu$ or $e$)~\cite{Aaij:2014ora,Aaij:2017vbb}.

Evidence for charged lepton universality violation is also observed in the charge current process $b\rightarrow c\tau\bar{\nu}$. The experiments, BaBar and Belle at the B-factories, made precise measurements of the ratios~\cite{Lees:2012xj,Lees:2013uzd,Huschle:2015rga,Sato:2016svk,Hirose:2016wfn}
\begin{equation}
R_{D} = \frac{\Gamma(B\rightarrow D\,\tau\,\bar{\nu})} {\Gamma(B\rightarrow D\, \{e/\mu\} \, \bar{\nu})},\hspace{0.1cm}
R_{D^*} = \frac{\Gamma(B\rightarrow D^{*}\,\tau\,\bar{\nu})} {\Gamma(B\rightarrow D^{*}\, \{e/\mu\} \, \bar{\nu})}.
\label{rdrds}
\end{equation}
These measurements are about $4\sigma$ away from the SM predictions. Very recently, LHCb experiment has measured $R_{D^*}$  and confirmed the discrepancy~\cite{Aaij:2015yra,Aaij:2017uff,Aaij:2017deq}. The measured experimental average values and the SM predictions for these ratios are given in table~\ref{tab1}.

%%%%%%%%%%%%%%%%%%%%%%%%%%%%%%%
\begin{table}[htbp]
\centering 
\tabcolsep 6pt
   
\begin{tabular}{cccc}
\hline\hline
      &  $R_D$  & $R_{D^*}$  & Correlation\\
\hline
Experimental average &  $0.407\pm 0.039\pm 0.024$ & $	0.304\pm 0.013\pm 0.007$ & $-0.20$ \\
 \hline
SM prediction  &  $0.300\pm 0.008$ & $0.252\pm 0.003$ & $-$\\
\hline\hline
\end{tabular}

\caption{Current world average of $R_D$/$R_{D^*}$~\cite{average} and their SM predictions  for $R_D$~\cite{Aoki:2016frl} and  $R_{D^*}$~\cite{Fajfer:2012vx}. The first (second) experimental errors are statistical (systematic).}
\label{tab1}
\end{table}
%%%%%%%%%%%%%%%%%%%%%%%%%%%%%%%

Recently  several groups have  updated the theoretical predictions of $R_D/R_{D^*}$ using different approaches, see for e.g., refs. ~\cite{Bigi:2016mdz,Bernlochner:2017jka,Bigi:2017jbd,Jaiswal:2017rve}. Ref. \cite{Bigi:2016mdz} improved the SM prediction of $R_{D}$ by making use of the lattice calculations of $B\rightarrow D\, l\,  \bar{\nu}$ form factors \cite{Lattice:2015rga,Na:2015kha} along with stronger unitarity constraints. It is the most precise prediction for $R_{D}$ till date. The value of $R_{D^{*}}$ has been updated in  \cite{Bernlochner:2017jka} by performing a combined fit to the  $B\rightarrow D^{(*)}\, l\,  \bar{\nu}$ decay distributions and including uncertainties in the form factor ratios at ${\mathcal{O}}(\alpha_s,\,\Lambda_{\rm QCD}/m_{c,b})$ in Heavy Quark Effective Theory (HQET). Ref. \cite{Bigi:2017jbd} obtained the SM prediction for $R_{D^{*}}$ by using heavy quark
symmetry relations between the form factors and including recent inputs from lattice calculations and experiments. The SM prediction for $R_{D^{*}}$ was obtained in \cite{Jaiswal:2017rve} by including the available known corrections at ${\mathcal{O}}(\alpha_s,\,\Lambda_{\rm QCD}/m_{c,b})$
in the HQET relations between the form factors along with the unknown
corrections in the ratios of the HQET form factors. This is done by introducing additional factors
and fitting them from the experimental data and lattice inputs.

All the meson decays in eq.~\ref{rdrds} are driven by quark level transitions $b\rightarrow cl\bar{\nu}$. These transitions occur at tree level in the SM unlike the FCNI. The discrepancy between the measured values of $R_D$ and $R_{D^*}$ and their respective SM predictions is an indication of presence of new physics (NP) in the $b\rightarrow c\tau\bar{\nu}$ transition. The possibility of NP in $b\rightarrow c\mu\bar{\nu}$ is excluded by other data~\cite{Alok:2017qsi}. All possible NP four-Fermi interaction terms for $b\rightarrow c\tau\bar{\nu}$ transition are listed in ref.~\cite{Freytsis:2015qca}. In ref~\cite{Alok:2017qsi}, a fit was performed between all the $B\rightarrow D/D^*\tau\bar{\nu}$ data and each of the NP interaction term. The NP terms, which can account for the $R_D$/$R_{D^*}$ data and are consistent with the constraint from $B_c\rightarrow \tau\bar{\nu}$, are identified and their Wilson coefficients (WCs) are calculated. It was found that four distinct solutions, each with a different Lorentz structure, are allowed. 

In ref.~\cite{Alok:2016qyh}, an attempt was made to distinguish between the allowed solutions by means of $\langle f_L\rangle$, the $D^*$ polarization fraction. It was found that the NP solution with the tensor Lorentz structure could be distinguished from other possibilities provided $\langle f_L\rangle$ can be measured with an absolute uncertainty of $0.1$. It was also shown in refs.~\cite{Hirose:2016wfn,Alonso:2016oyd,Altmannshofer:2017poe} that the $\tau$ polarization fraction, $P^{D^*}_{\tau}$, in $B\rightarrow D^*\tau\bar{\nu}$ is also effective in discriminating NP tensor operator.
 Therefore, in order to uniquely determine the Lorentz structure of new physics in $b\rightarrow cl\bar{\nu}$,  one needs additional observables. 

In this letter, we consider the angular observables $A_{FB}$ (the forward-backward asymmetry) and $A_{LT}$ (longitudinal-transverse asymmetry) in the decay $B\rightarrow D^*\tau\bar{\nu}$, in addition to the $D^*$ and $\tau$ polarizations mentioned above. These asymmetries can only be measured if the momentum of the $\tau$ lepton is reconstructed. We will show below that a measurement of these asymmetries, together with $\tau$ and $D^*$ polarization, can uniquely identify the Lorentz structure of the NP operator responsible for the present discrepancy in $R_D$ and $R_{D^*}$, if each observable is measured to the desired accuracy.  

%%%%%%%%%%%%%%%%%%%%%%%%%%%%%%%%%%
\section{New physics solutions}
%%%%%%%%%%%%%%%%%%%%%%%%%%%%%
The most general effective Hamiltonian for $b\rightarrow c\tau\bar{\nu}$ transition can be written as \cite{Freytsis:2015qca}
\begin{equation}
H_{eff} = \frac{4 G_F}{\sqrt{2}} V_{cb} \left[O_{V_L} + \frac{\sqrt{2}}{4 G_F V_{cb}\Lambda^2}  \sum_i C^{(','')}_iO^{(','')}_i\right]
%= \frac{4 G_F}{\sqrt{2}} V_{cb} \left[O_{V_L} + \alpha  \sum_i C^{(','')}_iO^{(','')}_i\right]\,,
\label{effH}
\end{equation}
where $G_F$ is the Fermi coupling constant, $V_{cb}$ is the Cabibbo-Kobayashi-Maskawa (CKM) 
matrix element and the NP scale $\Lambda$ is assumed to be 1 TeV.
In eq.~\ref{effH}, the unprimed operators $O_i$ are given by,
\begin{equation}
O_{V_L} =(\bar c \gamma_{\mu} P_L b)(\bar \tau \gamma^{\mu}  P_L \nu) \ , \quad  
O_{V_R}=(\bar c \gamma_{\mu}  P_R b)(\bar \tau \gamma^{\mu}  P_L \nu) \ , \quad  \nonumber
\end{equation}
\begin{equation}
O_{S_R}=(\bar c P_R b)(\bar \tau P_L \nu)  \ , \quad  
O_{S_L}=(\bar c P_L b)(\bar \tau P_L \nu ) \ , \quad   \nonumber
\end{equation}
\begin{equation}
O_T=(\bar c \sigma_{\mu \nu}P_L b)(\bar \tau \sigma^{\mu \nu} P_L \nu) \ . \quad
\label{ops}
\end{equation}
We also assume that neutrino is always left chiral. The effective Hamiltonian for the SM contains only the $O_{V_L}$ operator. The NP operators $\mathcal{O}_i$, $\mathcal{O}^{'}_i$ 
and $\mathcal{O}^{''}_i$ in the low energy effective Hamiltonian include all other 
possible Lorentz structures. The NP effects are encoded in the NP WCs $C_i, C^{'}_i$ and $C^{''}_i$.
The primed operators $O'_i$ are products of lepton-quark bilinears $\bar{\tau} \Gamma b$ and $\bar{c}\Gamma \nu$, where $\Gamma$ is a generic Dirac matrix. The double primed operators $O''_i$ couple the bilinear of 
form $\bar{\tau} \Gamma c^c$ to $\bar{b^c}\Gamma \nu$. 
Through Feirz transformation, each primed and double primed operator can be expressed as a linear combination of unprimed operators~\cite{Freytsis:2015qca}.

In an earlier report, we have calculated the values of NP WCs which fit the data on the observables $R_D$, $R_{D^*}$, $R_{J/\psi}$, $\langle P_{\tau}^{D^*}\rangle$ and $\mathcal{B}(B_c\rightarrow \tau\bar{\nu})$~\cite{Alok:2017qsi}. Here $R_{J/\psi}$ is the ratio of $\mathcal{B}(B_c\rightarrow J/\psi\tau\bar{\nu})$ to $\mathcal{B}(B_c\rightarrow J/\psi\mu\bar{\nu})$~\cite{Aaij:2017tyk}. In doing these calculations we have considered either one NP operator at a time or two similar operators at a time, such as ($O_{V_L}$, $O_{V_R}$) and ($O''_{S_L}$, $O''_{S_R}$). The results of these fits are listed in table~\ref{tab2}. This table also lists, for each of the NP solutions, the predicted values of the polarization fractions and the angular asymmetries in $B\rightarrow D^*\tau\bar{\nu}$ decay.
%%%%%%%%%%%%%%%%
\begin{table}[htbp]
\centering 
\tabcolsep 6pt
\begin{tabular}{|c|c|c|c|c|c|}
\hline\hline
   NP WCs  & Fit values  &$\langle P^{D^*}_{\tau}\rangle$ & $\langle f_L\rangle$&$\langle A_{FB}\rangle$  & $\langle A_{LT}\rangle$ \\
\hline
SM  &$C_{i}=0$   &$-0.499\pm 0.004$ &$0.45\pm 0.04$& $-0.011\pm 0.007$ & $-0.245\pm 0.003$ \\
\hline 
$C_{V_L}$  & $0.149\pm 0.032$ &$-0.499\pm 0.004$&  $0.45\pm 0.04$ &$-0.011\pm 0.007$ &$-0.245\pm 0.003$  \\

\hline
$C_T$  &  $0.516 \pm 0.015$&$+0.115\pm 0.013$ & $0.14\pm 0.03$  &$-0.114\pm 0.009$& $+0.110\pm 0.009$ \\

\hline
$C''_{S_L}$ & $-0.526\pm 0.102$  &$-0.485\pm 0.003$&$0.46\pm 0.04$ &$-0.087\pm 0.011$&$-0.211\pm 0.008$ \\

\hline
$(C_{V_L},C_{V_R})$& $(-1.286, 1.512)$ & $-0.499\pm 0.004$& $0.45\pm 0.04$&$-0.371\pm 0.004$&$+0.007\pm 0.004$ \\

\hline
$(C'_{V_L},\, C'_{V_R})$  &  $(0.124, -0.058)$ & $-0.484\pm 0.005$ &$0.45\pm 0.04$& $-0.003\pm 0.007$ & $-0.243\pm 0.003$  \\

\hline
$(C''_{S_L},\, C''_{S_R})$  & $(-0.643, -0.076)$ & $-0.477\pm 0.003$&$0.46\pm 0.04$& $-0.104\pm 0.005$& $-0.202\pm 0.002$  \\
\hline\hline
\end{tabular}

\caption{Best fit values of NP WCs at $\Lambda=1$ TeV, taken from table IV of ref.~\cite{Alok:2017qsi}. We provide the predictions of  $\langle P^{D^*}_{\tau}\rangle$, $\langle f_L\rangle$, $\langle A_{FB}\rangle$ and $\langle A_{LT}\rangle$ in decay $B\rightarrow D^*\tau\bar{\nu}$ with their uncertainties for each of the allowed solutions.  }
\label{tab2}
\end{table}
%%%%%%%%%%%%%%%%%%%%%%%
These observables are standard tools to discriminate between terms in an effective Hamiltonian with different Lorentz structures~\cite{Alok:2010zd,Celis:2012dk,Becirevic:2016hea,Bardhan:2016uhr,Alonso:2017ktd,Jung:2018lfu}. Here we compute $P_{\tau}^{D^*}(q^2)$, $f_{L}(q^2)$, $A_{FB}(q^2)$ and $A_{LT}(q^2)$ in $B\rightarrow D^*\tau\bar{\nu}$ decay, as functions of $q^2 = (p_B - p_{D^*})^2$, where $p_B$ and $p_{D^*}$ are the four momenta of $B$ and $D^*$ respectively. These observables are defined as 
\begin{eqnarray}
P^{D^{*}}_{\tau} (q^2) &=& \frac{(d\Gamma/dq^2)_{\lambda_{\tau}=1/2}-(d\Gamma/dq^2)_{\lambda_{\tau}=-1/2}}{(d\Gamma/dq^2)_{\lambda_{\tau}=1/2}+(d\Gamma/dq^2)_{\lambda_{\tau}=-1/2}},
\label{ptau}\\
A_{FB} (q^2)&=& \frac{1}{d\Gamma /dq^2}\left[\int^1_0\frac{d^2\Gamma }{dq^2d\cos\theta_{\tau}}d\cos\theta_{\tau}-\int^0_{-1}\frac{d^2\Gamma }{dq^2d\cos\theta_{\tau}}d\cos\theta_{\tau}\right],
\label{afb}\\
f_L (q^2)& = & \frac{(d\Gamma/dq^2)_{\lambda_{D^*}=0}}{(d\Gamma/dq^2)_{\lambda_{D^*}=0}+(d\Gamma/dq^2)_{\lambda_{D^*}=-1}+(d\Gamma/dq^2)_{\lambda_{D^*}=+1}},
\label{fl}\\
A_{LT} (q^2) &=& \frac{\int^{\pi/2}_{-\pi/2}d\phi\left(\int^1_0 d\cos\theta_D\frac{d^3\Gamma}{dq^2d\phi\, d\cos\theta_D}-\int^0_{-1} d\cos\theta_D\frac{d^3\Gamma}{dq^2d\phi\, d\cos\theta_D}\right)}{\int^{\pi/2}_{-\pi/2}d\phi\left(\int^1_0 d\cos\theta_D\frac{d^3\Gamma}{dq^2d\phi\, d\cos\theta_D}+\int^0_{-1} d\cos\theta_D\frac{d^3\Gamma}{dq^2d\phi\, d\cos\theta_D}\right)}.
\label{alt}
\end{eqnarray} 
Here  $\theta_D$ is the angle between $B$ and $D$ mesons where $D$ meson comes from $D^*$ decay, $\theta_{\tau}$ is the angle between $\tau$ and $B$ and  $\phi$ is the angle between $D^*$ decay plane and the plane defined by the tau momenta. The predictions for $P_{\tau}^{D^*}(q^2)$, $f_{L}(q^2)$ and $A_{FB}(q^2)$ are calculated using the framework provided in \cite{Sakaki:2013bfa} and for $A_{LT}(q^2)$ we follow ref~\cite{Duraisamy:2014sna}.
We also analyze tau polarization and forward backward asymmetry in $B\rightarrow D\tau\bar{\nu}$ decay.   The definitions for these observables are similar to that of the correspoinding observables in $B\rightarrow D^* \tau\bar{\nu}$, defined in eqs. \ref{ptau} and \ref{afb}.  We follow the method of ref. \cite{Sakaki:2013bfa} in calculating these quantities. 

The  $B\rightarrow D^{(*)}\, l\,  \bar{\nu}$ decay distributions depend upon hadronic form-factors. So far, the determination of these form-factors depends heavily on HQET techniques.  In this work we use the HQET form factors, parametrized by Caprini {\it et al.} \cite{Caprini:1997mu}. The parameters for $B\rightarrow D$ decay are well known in lattice QCD \cite{Aoki:2016frl} and we use them in our analyses. For $B\rightarrow D^*$ decay, the HQET parameters are extracted using data from Belle and BaBar experiments along with lattice inputs. In this work, the numerical values of these parameters are taken from refs. \cite{Bailey:2014tva} and  \cite{Amhis:2016xyh}. The common normalization term of all the form factors, which is theoretically calculated in lattice~\cite{Bailey:2014tva}, cancels out in all the ratios defined in eqs.~(\ref{ptau})-(\ref{alt}). Hence all the inputs for our calculations are derived from fits to experiments within HQET framework.

This table lists six different NP solutions but only the first four solutions are distinct. The predictions for various observables for solution 6 are essentially equal to those for solution 3 because values of $C''_{S_L}$ for these two solutions are very close and the value of $C''_{S_R}$ in solution 6 is much smaller. Similarly we can argue that solution 5 is essentially equivalent to solution 1 because (a) Fierz transform of $O'_{V_L}$ is $O_{V_L}$, (b) value of $C'_{V_L}$ in solution 5 is close to the value of $C_{V_L}$ in solution 1 and (c) the value of $C'_{V_R}$ is smaller. Thus we have four different NP solutions with different Lorentz structures. We explore methods to distinguish between them.

%%%%%%%%%%%%%%%%%%%%%%%
\begin{figure}[t]
%[htbp] 
\centering
\begin{tabular}{ccc}
\includegraphics[width=64mm]{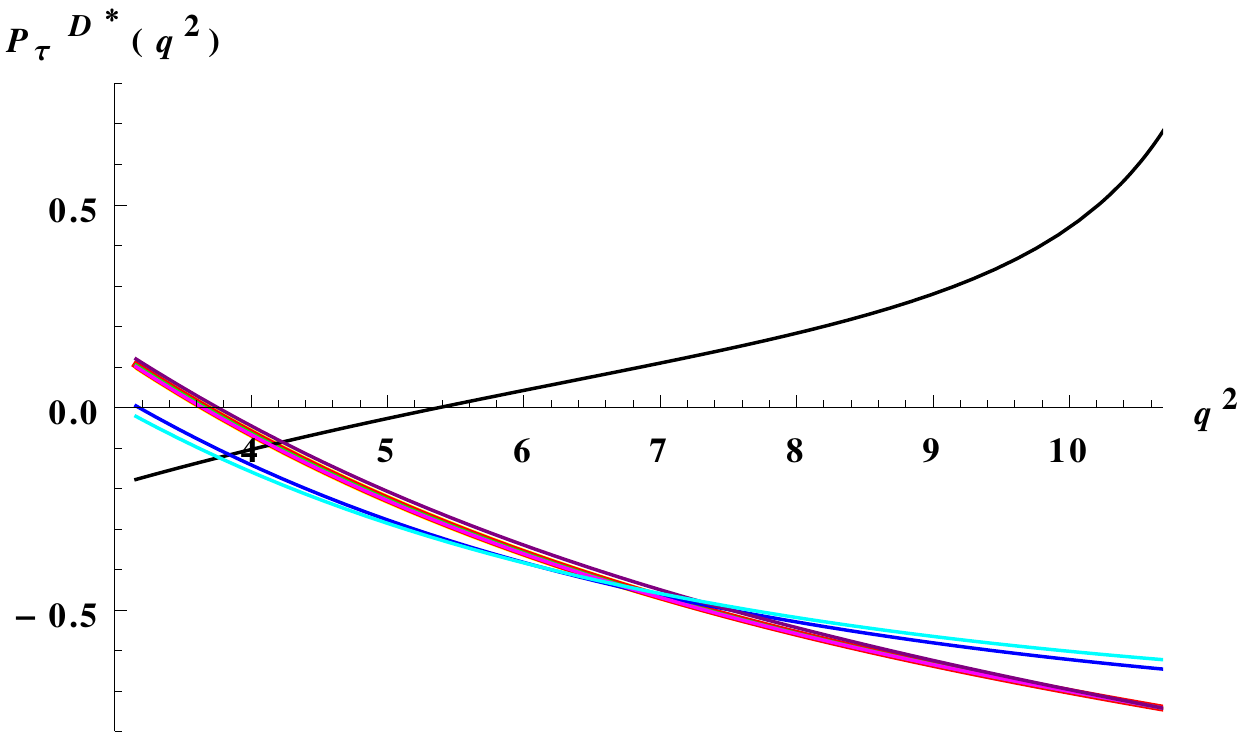}& \hspace{1cm} &
\includegraphics[width=60mm]{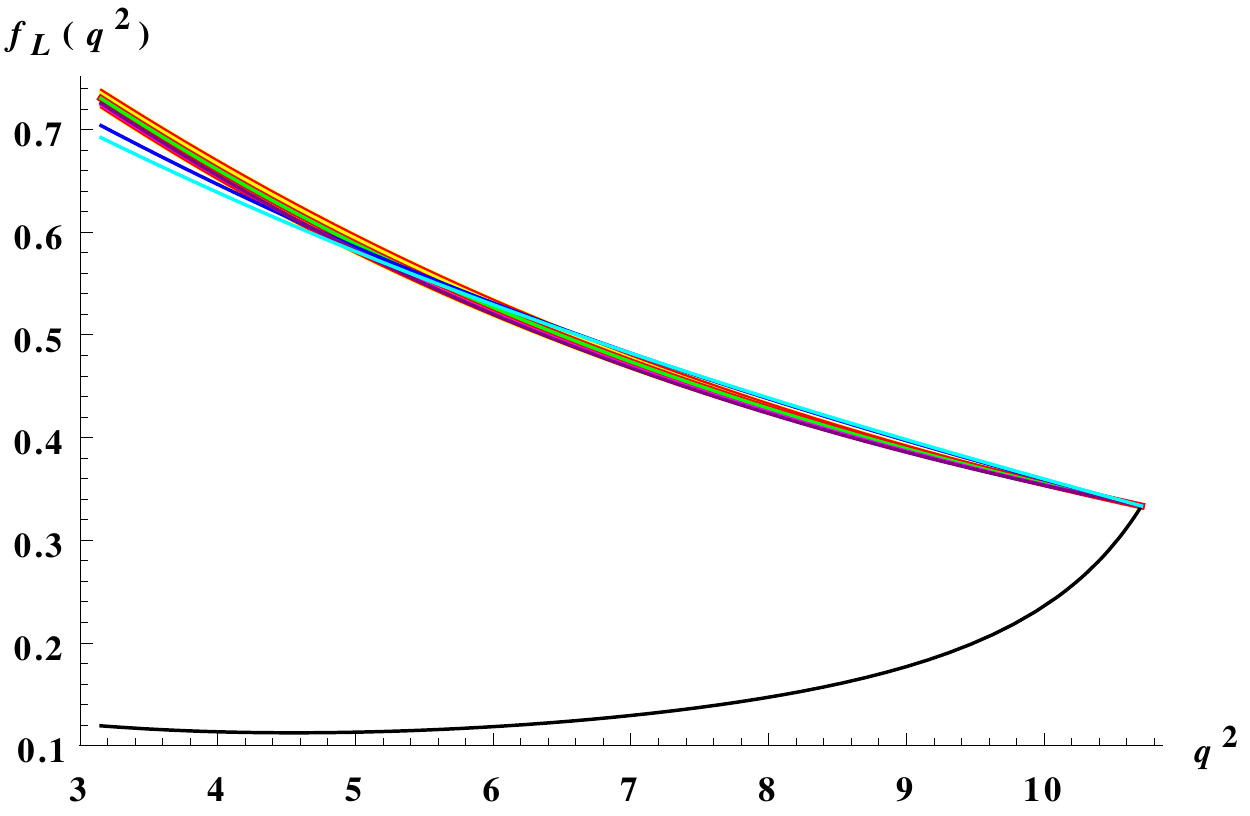}\\
\includegraphics[width=60mm]{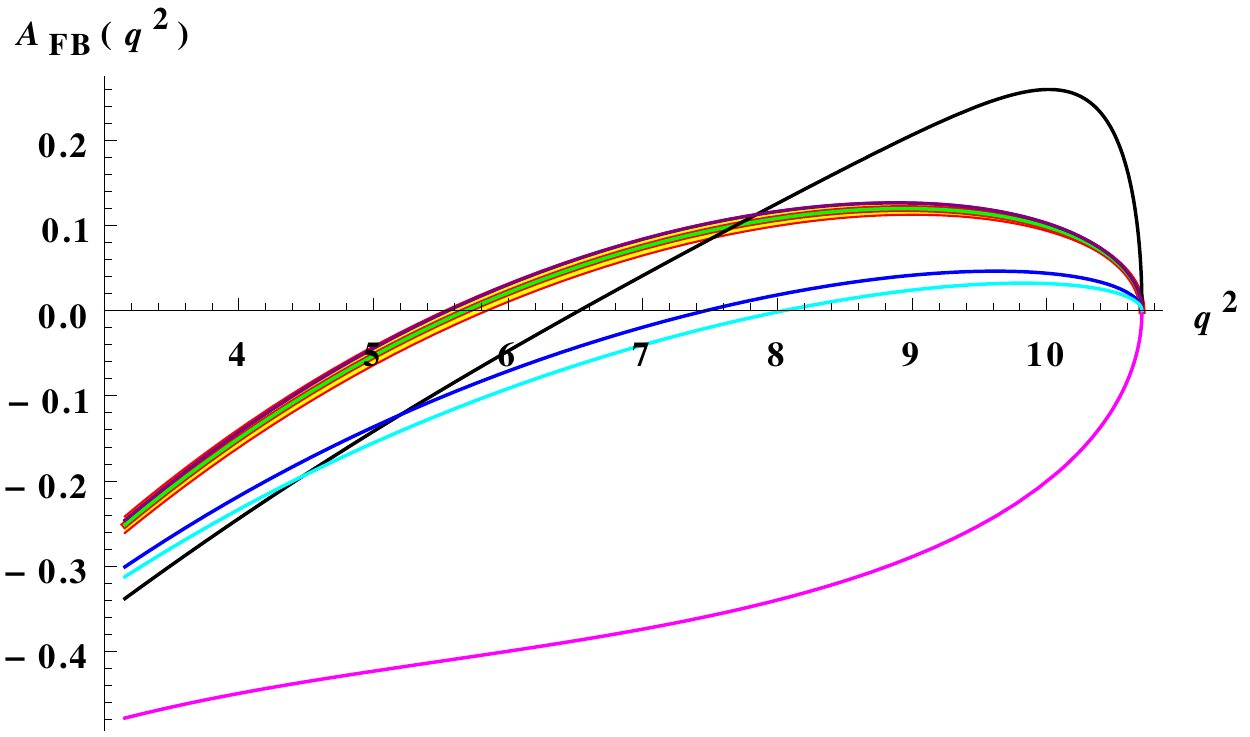}& \hspace{1cm} &
\includegraphics[width=60mm]{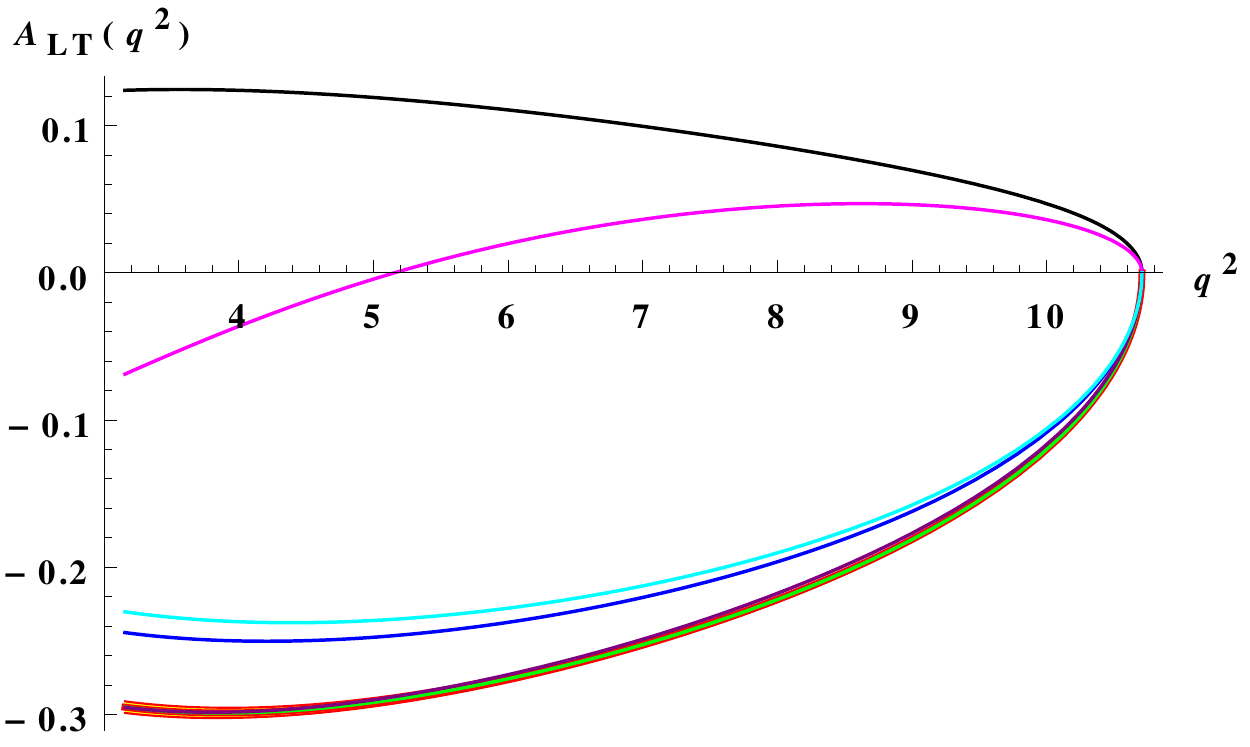} \\
\end{tabular}

\caption{Left and right panels in the top row correspond to $P^{D^*}_{\tau}(q^2)$ and $f_L(q^2)$, respectively for the $B\rightarrow D^*\tau\bar{\nu}$ decay whereas the left and right panels of bottom row correspond to $A_{FB}(q^2)$ and $A_{LT}(q^2)$. Red curves with yellow band corresponds to SM predictions. The band, representing 1$\sigma$ range, is mainly due to the uncertainties in various hadronic form factors and is obtained by adding these errors in quadrature. In each panel, the color code for the NP solutions is: $C_{V_L} = 0.149$ (green curve), $C_T = 0.516$ (black curve), $C''_{S_L} = -0.526$ (blue curve), $(C_{V_L},C_{V_R})= (-1.286,1.512)$ (magenta curve), $(C'_{V_L},C'_{V_R}) = (0.124, -0.058)$ (purple curve), $(C''_{S_L}, C''_{S_R}) = (-0.643, -0.076)$ (cyan curve).}
\label{fig1}
\end{figure}
%%%%%%%%%%%%%%%%%%%%%%%%%
%%%%%%%%%%%%%%%%%%%%%%%
\section{Results and discussion}
%%%%%%%%%%%%%%%%%%%%%%%%
The variation of $P_{\tau}^{D^*}$ and $f_{L}$ with $q^2$ is shown in the top row of fig.~\ref{fig1}. From these plots, we see that the plots of $O_T$ solution for both these variables differ significantly from the plots of other NP solutions. The average values of these observables, for each NP solution, are listed in table~\ref{tab2}. Not surprisingly, there is a large difference between the predicted values for $O_T$ solution and those for other NP solutions. If either of these observables is measured with an absolute uncertainty of $0.1$, then the $O_T$ solution is either confirmed or ruled out at $3\sigma$ level. It is interesting to note that the Belle collaboration has already made an effort to measure $\langle P_{\tau}^{D^*}\rangle$~\cite{Sato:2016svk} though the error bars are very large. They are also in the process of measuring $\langle f_L\rangle$~\cite{Adamczyk}.

Our ability to measure the angular observables $A_{FB}$ and $A_{LT}$ crucially depends on our ability to reconstruct the $\tau$ momentum. This may be very difficult to do because of the missing neutrino in the $\tau$ decay. However, as we will show below, these asymmetries are capable of distinguishing between the three remaining NP solutions. Hence it is imperative to develop methods to reconstruct the $\tau$ momentum.

The plots for $A_{FB}$ and $A_{LT}$ as a function of $q^2$ are shown in the bottom row of fig.~\ref{fig1} and their average values are listed in table~\ref{tab2}. We see that the plots of both $A_{FB}(q^2)$ and $A_{LT}(q^2)$, for $(O_{V_L}, O_{V_R})$ solution, differ significantly from the plots of all other NP solutions as do the average values. If either of these asymmetries is measured with an absolute uncertainty of $0.07$, then the $(O_{V_L}, O_{V_R})$ solution is either confirmed or ruled out at $3\sigma$ level.

%%%%%%%%%%%%%%%%%%%%%%%%%%%%%%%%%
\begin{figure}[htbp] 
\centering
\begin{tabular}{cc}
\includegraphics[width=60mm]{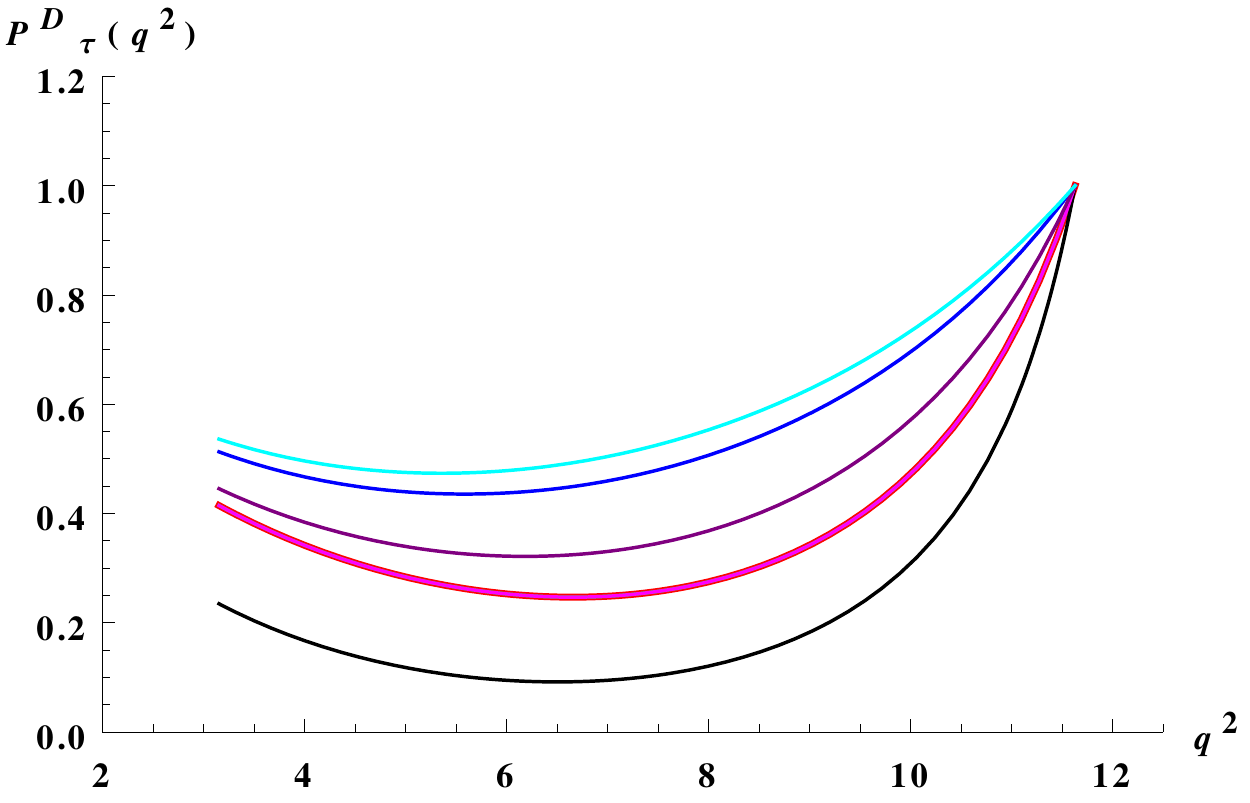}& 
\includegraphics[width=60mm]{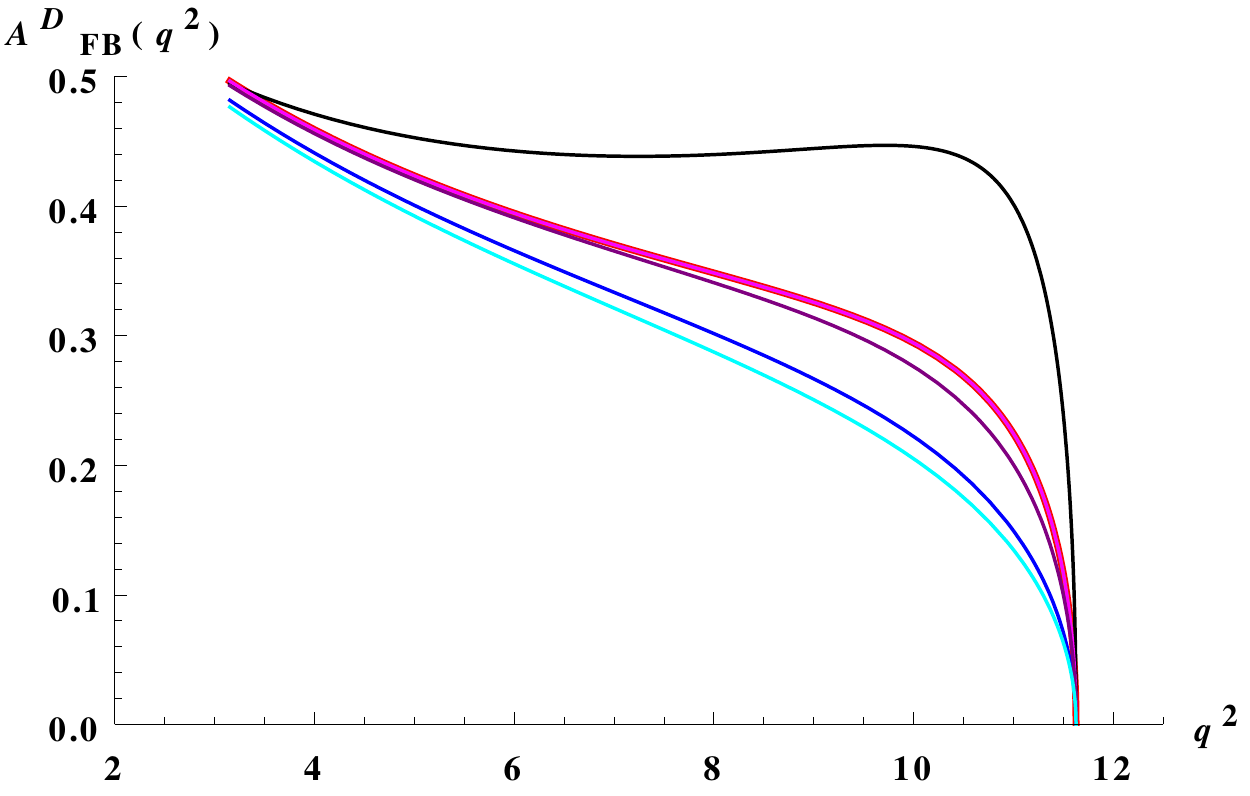} \\
\end{tabular}
\caption{Left and right panels correspond to $P^D_{\tau}(q^2)$ and $A^D_{FB}(q^2)$ in $B\rightarrow D\tau\bar{\nu}$ decay. Red curves with yellow band corresponds to SM predictions. The band  is obtained by adding errors, mainly due to hadronic form factors, in quadrature. $C_{V_L} = 0.149$ (green curve), $C_T = 0.516$ (black curve), $C''_{S_L} = -0.526$ (blue curve), $(C_{V_L},C_{V_R})= (-1.286,1.512)$ (magenta curve), $(C'_{V_L},C'_{V_R}) = (0.124, -0.058)$ (purple curve), $(C''_{S_L}, C''_{S_R}) = (-0.643, -0.076)$ (cyan curve) for each plot.}
\label{fig3}
\end{figure}
%%%%%%%%%%%%%%%%%%%%%%%%%%%%%%%%%

So far we have identified observables which can clearly identify the $O_T$ and the $(O_{V_L}, O_{V_R})$ solutions. As we can see from table~\ref{tab2}, one needs to measure $\langle A_{FB}\rangle$ with an absolute uncertainty of $0.03$ or better to obtain a $3\sigma$ distinction between $O_{V_L}$ and $O''_{S_L}$ solutions. However, this ability to make the distinction can be improved by observing $q^2$ dependence of $A_{FB}$ for these solutions. We note that $A_{FB}(q^2)$ for $O_{V_L}$ solution has a zero crossing at $q^2 = 5.6$ GeV$^2$ whereas this crossing point occurs at $q^2 = 7.5$ GeV$^2$ for $O''_{S_L}$ solution. A calculation of $\langle A_{FB}\rangle$ in the limited range $6$ GeV$^2< q^2<q^2_{max}$ gives the result $+0.1$ for $O_{V_L}$ and $+0.01$ for $O''_{S_L}$. Hence, determining the sign of $\langle A_{FB}\rangle$, for the full $q^2$ range and for the limited higher $q^2$ range, provides a very useful tool for discrimination between these two solutions.

%%%%%%%%%%%%%%%
\begin{table}[h]
\centering 
\tabcolsep 4pt
\begin{tabular}{|c|c|c|c|}
\hline\hline
   NP type  & Fit values &$\langle P^D_{\tau}\rangle$ &$\langle A^{D}_{FB}\rangle$  \\
\hline
SM  &$C_{i}=0$  &$0.325\pm 0.001$ & $0.360\pm 0.002 $\\
\hline  
$C_{V_L}$  & $0.149\pm 0.032$ &$0.325\pm 0.001$ & $0.360\pm 0.002$ \\
\hline
$C_T$  &  $0.516 \pm 0.015$ &$0.161\pm 0.001$ &$0.442\pm 0.002$ \\
\hline
$C''_{S_L}$ & $-0.526\pm 0.102$ & $0.538\pm 0.002 $  &$0.308\pm 0.002$\\
\hline
$(C_{V_L},C_{V_R})$& $(-1.286, 1.512)$&$0.325\pm 0.001$  &$0.360\pm 0.001$ \\
\hline
$(C'_{V_L},\, C'_{V_R})$  &  $(0.124, -0.058)$ &$0.410\pm 0.002$  & $0.348\pm 0.001$\\
\hline
$(C''_{S_L},\, C''_{S_R})$  & $(-0.643, -0.076)$ &$0.582\pm 0.002$ &$0.293\pm 0.001$  \\
\hline\hline
\end{tabular}
\caption{  Predictions of $\langle P^D_{\tau}\rangle$  and $\langle A^{D}_{FB}\rangle$ for $B\rightarrow D\tau\bar{\nu}$ decay.  }
\label{tab4}
\end{table}
%%%%%%%%%%%%%%%%%%%%%%%%%%%%

In principle the $\tau$ polarization and the forward backward asymmetry can be measured in $B\rightarrow D\tau\bar{\nu}$ decay also. The plots of $P^D_{\tau}(q^2)$ and $A^D_{FB}(q^2)$ vs. $q^2$ are given in fig.~\ref{fig3} and the average values are listed in table~\ref{tab4}. From this figure we see that only the plots for $O_T$ significantly differs from others, hence these observables have only a limited discriminating power.

\section{Conclusions}
In conclusion, we find that a clear distinction can be made between the four different NP solutions to the $R_D$/$R_{D^*}$ puzzle by means of polarization fractions and angular asymmetries. A measurement of either $\tau$ polarization or $D^*$ polarization with an absolute uncertainty of $0.1$ either confirms the $O_T$ solution as the explanation of the puzzle or rules it out. Similarly, the $(O_{V_L}, O_{V_R})$ solution is either confirmed or ruled out if one of the angular asymmetries, $\langle A_{FB}\rangle$ or $\langle A_{LT}\rangle$, is measured with an absolute uncertainty of $0.07$. Separating the $O_{V_L}$ and the $O''_{S_L}$ solutions is a little more difficult. But determining the sign of $\langle A_{FB}\rangle$ in the reduced  $q^2$ range (6 GeV$^2$, $q^2_{max}$) can lead to an additional distinction between these solutions provided a measurement of this asymmetry at the level $\approx 0.1$ is possible. Note that only the observables isolating $O_T$ do not require the reconstruction of $\tau$ momentum. This reconstruction of $\tau$ momentum is crucial to measure the asymmetries which can distinguish between the other three NP solutions. It is worth taking up this daunting challenge to clearly identify the type of NP which can explain the $R_D$/$R_{D^*}$ puzzle. 

\subsection*{Acknowledgements}
SUS thanks the theory group at CERN for their hospitality during the time
the manuscript of this letter is finalized. He also thanks Concezio Bozzi
and Greg Ciezarek for valuable discussions.

%%%%%%%%%%%%%%%%%%%%%%%%%%%%%%%%%%%%%%%%%%%%%%%

%%%%%%%%%%%%%%%%%%%%%%%%%%%%%%%%%%%%%%%%%%%%%%%%%5

\end{document}